\documentclass[aps,prb,twocolumn,superscriptaddress]{revtex4-1}
\usepackage{graphicx}

\begin{document}

\title{Magnon spectra and strong spin-lattice coupling in magnetically frustrated Mn$B_{2}$O$_{4}$ ($B$ = Mn,V): Inelastic light scattering studies}

\author{S.~L.~Gleason}
\thanks{These authors contributed equally to this work.}
\author{T.~Byrum}
\thanks{These authors contributed equally to this work.}
\author{Y.~Gim}
\author{A.~Thaler}
\author{P.~Abbamonte}
\author{G.~J.~MacDougall}
\affiliation{Department of Physics and Materials Research Laboratory, 
University of Illinois, Urbana, Illinois  61801, USA}

\author{L.~W.~Martin}
\affiliation{Department of Materials Science and Engineering and Materials Research Laboratory, University of Illinois, Urbana, Illinois  61801, USA}

\author{H.~D.~Zhou}
\affiliation{Department of Physics and Astronomy, University of Tennessee, Knoxville, Tennessee 37996-1200, USA}
\affiliation{National High Magnetic Field Laboratory, Florida State University, Tallahassee, Florida 32306-4005, USA}

\author{S.~L.~Cooper}
\affiliation{Department of Physics and Materials Research Laboratory, University of Illinois, Urbana, Illinois  61801, USA}

\date{\today}

\begin{abstract}

The ferrimagnetic spinels Mn$B_{2}$O$_{4}$ ($B$ = Mn,V) exhibit a similar series of closely spaced magnetic and structural phase transitions at low temperatures, reflecting both magnetic frustration and a strong coupling between the spin and lattice degrees of freedom.
Careful studies of excitations in Mn$B_{2}$O$_{4}$ ($B$ = Mn,V), and the evolution of these excitations with temperature, are important for obtaining a microscopic description of the role that magnetic excitations and spin-lattice coupling play in the low temperature phase transitions of these materials.
We report an inelastic light (Raman) scattering study of the temperature and magnetic field dependences of one- and two-magnon excitations in MnV$_{2}$O$_{4}$ and Mn$_{3}$O$_{4}$.
We observe a pair of $\textbf{q}=0$ one-magnon modes at 74~cm$^{-1}$ and 81~cm$^{-1}$ in MnV$_{2}$O$_{4}$, which is in contrast with the single 80~cm$^{-1}$ $\textbf{q}=0$ magnon that has been reported for MnV$_{2}$O$_{4}$ based on previous neutron scattering measurements and spin wave calculations.
Additionally, we find that the two-magnon energy of MnV$_{2}$O$_{4}$ decreases (``softens'') with decreasing temperature below $T_{N}$, which we attribute to strong coupling between magnetic and vibrational excitations near the zone boundary.

\end{abstract}


\maketitle

\section{Introduction}

Spinel compounds (chemical formula $AB_{2}X_{4}$) consist of an \textit{A}-site diamond sublattice and a geometrically frustrated \textit{B}-site pyrochlore sublattice.\cite{Lee2010}
A remarkable feature of these materials is the diversity of phenomena that result from the substitution of magnetic ions on the frustrated \textit{B}-site sublattice. 
For example, superconductivity,\cite{McCallum1976} charge ordering,\cite{Irizawa2011} and heavy fermion behavior\cite{Kopec2009} have all been reported in the Li$B_{2}$O$_{4}$ ($B$ = Ti,Mn,V) family of spinels. 
The ferrimagnetic spinels Mn$_{3}$O$_{4}$\cite{Jensen1974,Chardon1986,Chung2008} and MnV$_{2}$O$_{4}$\cite{Zhou2007,Suzuki2007,Garlea2008,Chung2008} offer a particularly interesting comparison:
Both systems exhibit a similar series of closely spaced magnetic and structural phase transitions at low temperatures, reflecting both magnetic frustration and a strong coupling between the spin and lattice degrees of freedom.
For example, MnV$_{2}$O$_{4}$ orders ferrimagnetically below $T_{N}=56 \textrm{ K}$, but transitions to a noncollinear ferrimagnetic phase following a cubic-to-tetragonal structural transition at $T_{o} = 53 \textrm{ K}$.\cite{Zhou2007,Suzuki2007,Garlea2008,Chung2008} 
Similarly, Mn$_{3}$O$_{4}$ orders in a noncollinear ferrimagnetic phase below $T_{N} = 42 \textrm{ K}$ before subsequently transitioning first to an incommensurate magnetic phase at $T_{1} = 39 \textrm{ K}$, and then to a cell-doubled version of the noncollinear ferrimagnetic phase below a tetragonal-to-orthorhombic structural transition at $T_{2} = 33 \textrm{ K}$.\cite{Jensen1974,Chardon1986} 
It remains an open question how these complex magnetostructural transitions in Mn$B_{2}$O$_{4}$ ($B$=Mn,V) are influenced by the orbital configurations, which are quite different in Mn$_{3}$O$_{4}$ and MnV$_{2}$O$_{4}$. 
While the \textit{B}-site Mn$^{3+}$ ion in Mn$_{3}$O$_{4}$ is locked into a specific orbital configuration
\footnote{
Specifically, in the high temperature cubic phase ($T > 1443 \textrm{ K}$), the Mn$^{3+}$ ($3d^{4}$) ions on the \textit{B}-sites of Mn$_{3}$O$_{4}$ have filled t$_{2g}$ levels and a single electron occupying the two-fold degenerate e$_{g}$ levels.  
However, the twofold orbital degeneracy of the e$_{g}$ states is lifted by a cooperative Jahn-Teller distortion involving a \textit{c}-axis lattice expansion at $T_{JT} = 1443 \textrm{ K}$, which is associated with $d_{3z^{2}-r^{2}}$ ferro-orbital order.}
below the cubic-to-tetragonal transition near $T_{o} = 1440 \textrm{ K}$,\cite{Jensen1974,Chardon1986,Chung2008,Nii2013} the \textit{B}-site V$^{3+}$ ion in MnV$_{2}$O$_{4}$ remains orbitally degenerate below the cubic-to-tetragonal structural transition at 53~K.\cite{Garlea2008,Chung2008,Huang2011} 
Various orbital configurations for the V$^{3+}$ ion in MnV$_{2}$O$_{4}$ have been proposed, including alternating $d_{xz}$ and $d_{yz}$ orbitals (\textit{L} = 0) along the \textit{c}-axis,\cite{Tsunetsugu2003} a complex $d_{xz}$+$id_{yz}$ (\textit{L} = 1) configuration,\cite{Nanguneri2012,Tchernyshyov2004} and even more complicated mixtures of $d$-orbital states.\cite{Khomskii2005,Chern2010,Katsufuji2013,Sarkar2009}

Careful studies of excitations in Mn$B_{2}$O$_{4}$ ($B$=Mn,V), and the evolution of these excitations with temperature, are important for obtaining a microscopic description of the role that magnetic excitations and spin-lattice coupling play in the low temperature phase transitions of these materials. 
To date, however, there has been little study of the temperature dependent evolution of the magnetic excitations in MnV$_{2}$O$_{4}$ and Mn$_{3}$O$_{4}$. 
In this paper, we report an inelastic light (Raman) scattering study of the temperature and magnetic field dependences of one- and two-magnon excitations in MnV$_{2}$O$_{4}$ and Mn$_{3}$O$_{4}$. 
We show that while the two-magnon energy of Mn$_{3}$O$_{4}$ has a conventional temperature dependence, the two-magnon energy of MnV$_{2}$O$_{4}$ exhibits an anomalous decrease (``softening'') with decreasing temperature below $T_{N}$, which we attribute to strong spin-lattice coupling associated with zone boundary magnons in MnV$_{2}$O$_{4}$. 
In addition, our high-resolution study of MnV$_{2}$O$_{4}$ at $T = 3 \textrm{ K}$ reveals a $\textbf{q}=0$ one-magnon spectrum that differs in important respects from previous neutron scattering results and spin wave calculations.

\section{Experiment}

\subsection{Sample preparation}

Single crystal samples of MnV$_{2}$O$_{4}$ were grown at Florida State University using a traveling-solvent-floating-zone technique.\cite{Zhou2007} 
The feed and seed rods for the crystal growth were prepared by solid state reaction. 
Appropriate mixtures of MnO and V$_{2}$O$_{3}$ were ground together and pressed into 6~mm diameter, 60~mm long rods under 400~atm hydrostatic pressure, and then calcined in vacuum in a sealed quartz tube at 950~$^{\circ}$C for 12~hours. 
The crystal growth was carried out in argon gas in an NEC IR-heated image furnace equipped with two halogen lamps and double ellipsoidal mirrors. 
The growth was conducted with the feed and seed rods rotating in opposite directions at 25~rpm during crystal growth at a rate of 30~mm/hour. 
Because of the evaporation of V$_{2}$O$_{3}$ during the growth, extra V$_{2}$O$_{3}$ in the starting material and high growth speeds are critical for obtaining high quality samples.
The structural and magnetic properties of the resulting crystal are reported elsewhere.\cite{Zhou2007}
The crystal exhibits two magnetic transitions at 56~K and 52~K,\cite{Zhou2007} consistent with previously reported temperatures.\cite{Suzuki2007,Garlea2008,Chung2008} 

Single crystal samples of Mn$_{3}$O$_{4}$ were grown at the University of Illinois using a floating-zone technique.\cite{Kim2010,Kim2011} 
Commercially available fine Mn$_{3}$O$_{4}$ powder (manganese (II, III) oxide, $\sim$325 mesh, 97\%, Sigma-Aldrich Co.) was used as the starting material. 
The powder was formed into feed/seed rods with a diameter of 10~mm and a length of 150~mm, and was pressed at a hydrostatic pressure of 400~atm. 
The pressed rods were sintered at 1050~$^{\circ}$C for 5~hours with an argon gas flow of 0.5~L/min. 
Single crystal growth was performed using a four-ellipsoid-mirror furnace (Crystal Systems Inc. FZ-T-10000-H-VI-VP) equipped with four 1000~W halogen lamps. 
The growth used a feed/seed (upper/lower shaft) rotation rate of 35 rpm in opposite directions, a growth rate (mirror-stage moving rate) of 5 mm/hour, a feeding rate (upper-shaft moving rate) of 1~mm/hour, and a growth atmosphere of oxygen at a gas pressure of 1~atm.
The structural and magnetic properties of the resulting crystal are reported elsewhere.\cite{Kim2011a,Kim2010,Kim2011}
The crystal exhibits three magnetic transitions at 43~K, 39~K, and 33~K, consistent with previously reported temperatures.\cite{Jensen1974,Chardon1986} 

The crystallographic orientations of the samples used in this experiment were determined via x-ray diffraction performed at room temperature.

\subsection{\label{Experiment}Raman scattering measurements}

Raman scattering measurements were performed using the 647.1~nm excitation line from a Kr$^{+}$ laser. 
The incident laser power was limited to 10~mW, and was focused to a $\sim$50~$\mu$m diameter spot to minimize laser heating of the sample, which was estimated to be roughly 4~K. 
The scattered light from the samples was collected in a backscattering geometry, dispersed through a triple stage spectrometer, and then detected with a liquid-nitrogen-cooled CCD detector. 
The samples were inserted into a continuous He-flow cryostat, which was horizontally mounted in the open bore of a superconducting magnet, allowing Raman measurements while the temperature was varied in the range $T = 3-290 \textrm{ K}$ and the magnetic field was varied in the range $H = 0-8 \textrm{ T}$.

For the scattering experiments involving Mn$_{3}$O$_{4}$, the incident light was circularly polarized with wavevector $\textbf{k}$ parallel to the \textit{c}-axis.
No analyzer was employed, in order to couple to as many components of the Raman tensor as possible in this geometry.
The magnetic field used in the magnetic field dependent measurements was directed along the \textit{c}-axis, parallel to $\textbf{k}$. 
For the scattering experiments involving MnV$_{2}$O$_{4}$, two experimental geometries were used. 
In geometry 1, the incident light was circularly polarized with $\textbf{k}$ parallel to a cubic axis. 
No analyzer was employed, in order to couple to as many components of the Raman tensor as possible in this geometry.
The magnetic field used in the magnetic field dependent measurements was directed along the same cubic axis, parallel to $\textbf{k}$.  
In geometry 2, a 1~T magnetic field was applied to define a unique [001] crystallographic direction in the sample,
\footnote{The application of a $H = 1 \textrm{ T}$ magnetic field along a cubic axis restored rigorous selection rules for the high energy phonons, indicating that the crystal had a single domain with $\textbf{M}\parallel\textbf{H}$.
Recent experiments (see Refs.~\onlinecite{Baek2009,Magee2010,Suzuki2007}) support this conclusion.}
where [hkl] refer to tetragonal axes.
The incident light was linearly polarized along [11$\sqrt{2}$] with $\textbf{k} \parallel [1\bar{1}0]$.  
An analyzer was employed to match the scattered polarization with the incident polarization along [11$\sqrt{2}$].
This geometry was used to equalize the intensities of the one-magnon excitations observed in MnV$_2$O$_4$ (see Fig.~\ref{fig:MnV2O4fvT}~(d)).

\section{Results}

\begin{figure}[t]
\includegraphics{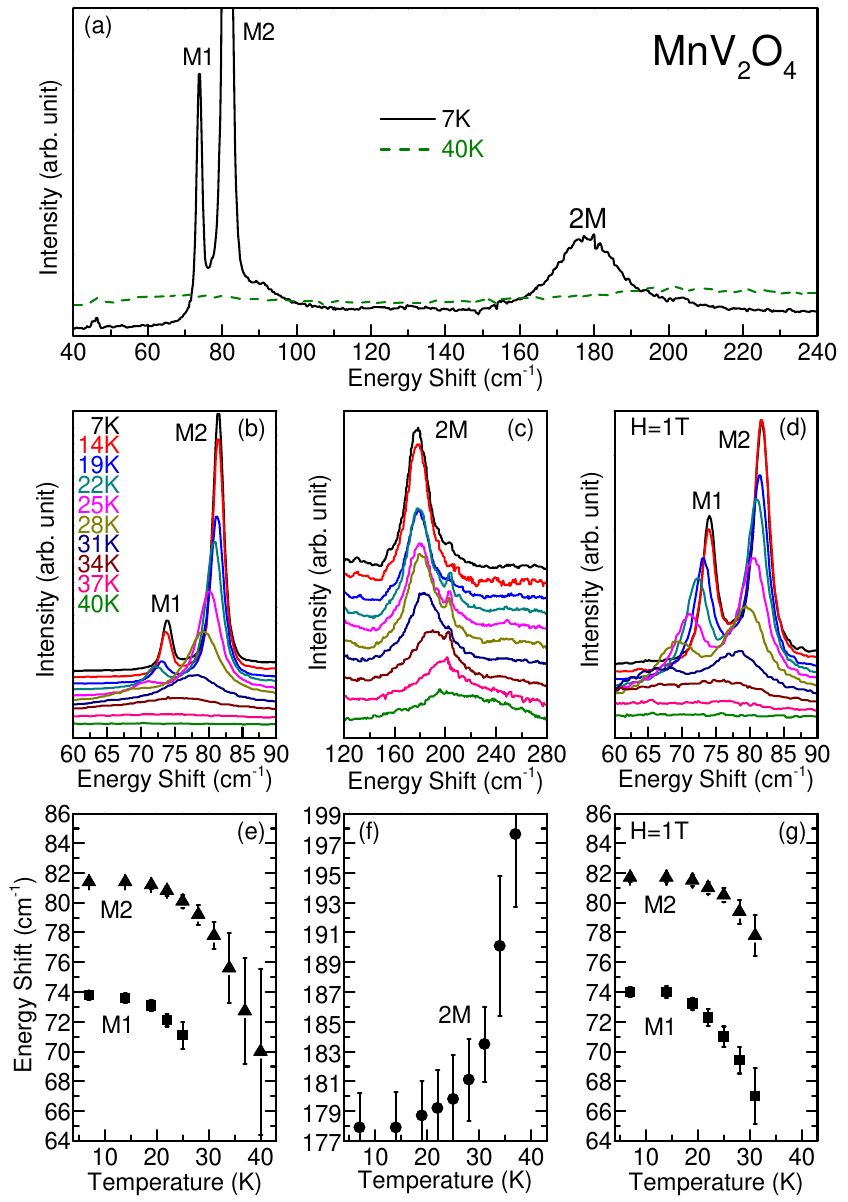}
\caption{\label{fig:MnV2O4fvT}
(color)
(a)-(c) Raman scattering spectra of MnV$_{2}$O$_{4}$ taken in geometry 1 (see discussion in \S{}\ref{Experiment}) at various temperatures in the energy ranges (a) $40-240 \textrm{ cm}^{-1}$, (b) $60-90 \textrm{ cm}^{-1}$, and (c) $120-280 \textrm{ cm}^{-1}$. 
(d) Raman scattering spectra of MnV$_{2}$O$_{4}$ taken in geometry 2 (see discussion in \S{}\ref{Experiment}) at various temperatures in the energy range $60-90 \textrm{ cm}^{-1}$. 
The data in (b)-(d) have been offset for clarity. 
(e)-(g) Summaries of the temperature dependences of the peak positions for peaks labeled M1, M2, and 2M, as estimated by eye.}
\end{figure}

Figure~\ref{fig:MnV2O4fvT} shows the temperature dependence of the MnV$_{2}$O$_{4}$ Raman spectrum in the $40-240 \textrm{ cm}^{-1}$ energy range, which is expected to be dominated by magnon excitations.\cite{Chung2008}
The higher frequency phonon spectrum we observe in MnV$_{2}$O$_{4}$ is consistent with that reported previously\cite{Takubo2011} and will not be reproduced here.
\footnote{The energies of the phonons we observe agree with those reported by Takubo \textit{et al}. 
Our symmetry assignments, however, differ from their assignments (see Ref. \onlinecite{Takubo2011}).}
For temperatures near and above $T_{N}$, the Raman spectrum of MnV$_{2}$O$_{4}$ exhibits a broad continuum background.
Since MnV$_{2}$O$_{4}$ is insulating,\cite{Nii2012} it is unlikely that this continuum background represents inelastic electronic scattering.
Rather, we associate this background with incoherent spin scattering similar to that observed in Raman scattering from low-dimensional spin systems\cite{Lemmens2003} and Cr-based spinels.\cite{Choi2007} 
For temperatures $T < T_{N}$, this incoherent spin scattering is suppressed, and several peaks develop, including two sharp modes at 74~cm$^{-1}$ (M1) and 81~cm$^{-1}$ (M2), a broad mode centered at 178~cm$^{-1}$ (2M), and a weak ``shoulder'' near 90~cm$^{-1}$ that will be discussed in more detail in a later publication.

\begin{figure}[t]
\includegraphics{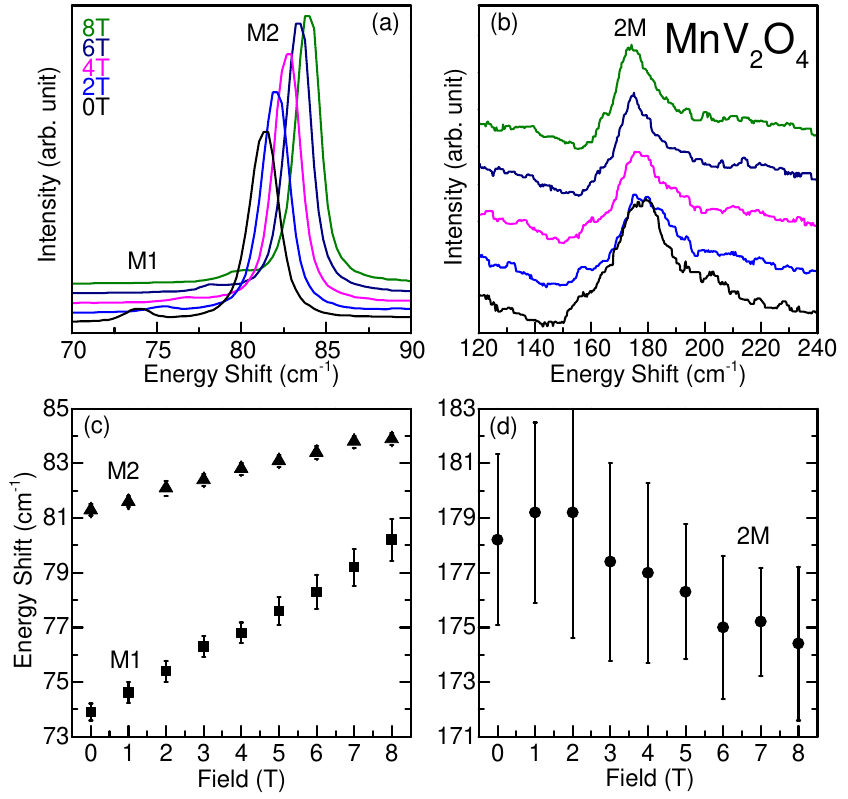}
\caption{\label{fig:MnV2O4fvH}
(color)
(a),(b) Raman scattering spectra of MnV$_{2}$O$_{4}$ taken in geometry 1 (see discussion in \S{}\ref{Experiment}) at various magnetic fields in the energy ranges (a) $70-90 \textrm{ cm}^{-1}$ and (b) $120-240 \textrm{ cm}^{-1}$. 
The data have been offset for clarity. 
(c),(d) Summaries of the magnetic field dependences of the peak positions for peaks labeled M1, M2, and 2M, as estimated by eye.}
\end{figure}

We identify the narrow peaks M1 and M2 as $\textbf{q}=0$ one-magnon excitations, based upon the following:
Both peaks broaden and decrease in energy as $T \rightarrow T_{N}$, tracking the sublattice magnetization, and are absent at temperatures above $T_{N}$. 
Additionally, as shown in Fig.~\ref{fig:MnV2O4fvH}, the energies of M1 and M2 change linearly with increasing magnetic field applied along the net magnetization direction --- with rates of $\Delta E/\Delta H = +0.75 \textrm{ cm}^{-1}$/T and $\Delta E/\Delta H = +0.33 \textrm{ cm}^{-1}$/T, respectively --- consistent with the field dependence expected of one-magnon excitations.\cite{Cottam1986} 
Finally, previous inelastic neutron scattering measurements of MnV$_{2}$O$_{4}$ confirm that there are magnon branches very close in energy to M1 (74~cm$^{-1}$) and M2 (81~cm$^{-1}$):  
One neutron study reports a $\textbf{q}=0$ one-magnon band near 80~cm$^{-1}$,\cite{Chung2008} while a second study reports the crossing of two one-magnon bands near 80~cm$^{-1}$ at $\textbf{q}=0$.\cite{Magee2010}

The broad band 2M in Fig.~\ref{fig:MnV2O4fvT} was previously observed in Raman scattering measurements and was attributed to one-magnon scattering.\cite{Takubo2011,Miyahara2010}
Indeed, a fit to inelastic neutron scattering data indicates that there are two one-magnon branches in the vicinity of 180~cm$^{-1}$ at $\textbf{q} = 0$.\cite{Chung2008} 
However, while we cannot rule out the possibility that the 2M band in MnV$_{2}$O$_{4}$ has contributions from one-magnon scattering, our discovery of one-magnon excitations M1 and M2 with linewidths an order of magnitude smaller than that of 2M makes it unlikely that the 2M band in Fig.~\ref{fig:MnV2O4fvT}~(a) can be attributed to a convolution of two closely spaced one-magnon peaks.
It is more likely that the 2M band is associated with two-magnon scattering, involving the excitation of magnon pairs with momenta +$\textbf{q}$ and $-\textbf{q}$. 
The substantially larger linewidth of 2M relative to M1 and M2 reflects the fact that the two-magnon scattering response is governed by the two-magnon density of states.

Interestingly, the temperature dependence of the two-magnon response in MnV$_{2}$O$_{4}$ is anomalous, as its energy \emph{increases} with increasing temperature towards $T_{N}$ (see Fig.~\ref{fig:MnV2O4fvT}). 
By contrast, the two-magnon scattering energy in magnetic materials is normally expected to decrease with increasing temperature towards $T_{N}$,\cite{Cottam1986} reflecting a decrease in the one-magnon energies and spin-spin correlations with increasing temperature. 
As discussed below, we propose that the anomalous temperature dependence of the two-magnon scattering response in MnV$_{2}$O$_{4}$ is indicative of strong spin-lattice coupling associated with zone boundary magnons.

\begin{figure}[t]
\includegraphics{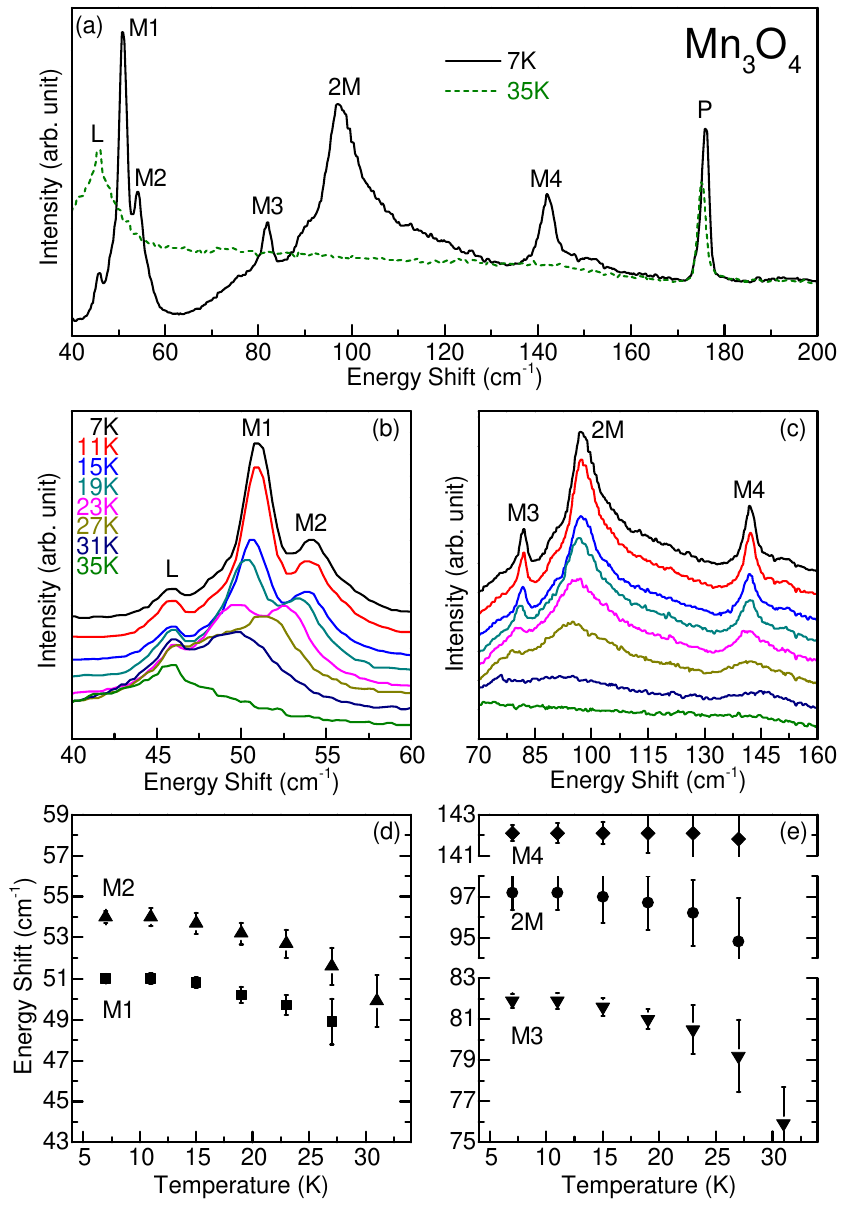}
\caption{\label{fig:Mn3O4fvT}
(color)
(a)-(c) Raman scattering spectra of Mn$_{3}$O$_{4}$ at various temperatures in the energy ranges (a) $40-200 \textrm{ cm}^{-1}$, (b) $40-60 \textrm{ cm}^{-1}$, and (c) $70-160 \textrm{ cm}^{-1}$. 
The data in (b) and (c) have been offset for clarity. 
(d),(e) Summaries of the temperature dependences of the peak positions for peaks labeled M1$-$M4 and 2M, as estimated by eye.}
\end{figure}

Figure~\ref{fig:Mn3O4fvT} shows the temperature dependence of the Mn$_{3}$O$_{4}$ Raman spectrum in the $40-200 \textrm{ cm}^{-1}$ energy range. 
The higher frequency Raman spectrum of Mn$_{3}$O$_{4}$, which is dominated by phonons, was reported by Kim \textit{et al.}\cite{Kim2010,Kim2011} and will not be reproduced or discussed here. 
The sharp peak labeled P persists to room temperature and is assigned to a phonon mode. 
For temperatures near and above $T_{N}$, the Raman spectrum of Mn$_{3}$O$_{4}$ exhibits a broad continuum background similar to that observed in MnV$_{2}$O$_{4}$. 
Again, as Mn$_{3}$O$_{4}$ is insulating,\cite{West1985} we attribute this background to inelastic magnetic, rather than electronic, scattering.
For temperatures $T < T_{N}$, the magnetic continuum background response is suppressed, and several peaks develop: 
Four narrow peaks at 51~cm$^{-1}$ (M1), 54~cm$^{-1}$ (M2), 82~cm$^{-1}$ (M3), and 142~cm$^{-1}$ (M4), and a broad, asymmetric band at 97.5~cm$^{-1}$ (2M). 
The weak feature labeled L is an artifact caused by stray light and will not be discussed further.
\footnote{The apparent broadening of L at 35~K is simply an artifact caused by the softening of magnetic excitations M1 and M2.}

We identify the four narrow peaks M1$-$M4 as $\textbf{q}=0$ one-magnon excitations, based upon the same criteria discussed above for MnV$_{2}$O$_{4}$: All of these peaks, except M4, broaden and decrease in energy as $T \rightarrow T_{N}$. 
Additionally, the energies of the $\textbf{q}=0$ one-magnon modes M1$-$M4 in Fig.~\ref{fig:Mn3O4fvT} correspond well with the $\textbf{q}=0$ magnon energies reported in a previous inelastic neutron scattering study.\cite{Chung2008}

\begin{figure}[t]
\includegraphics{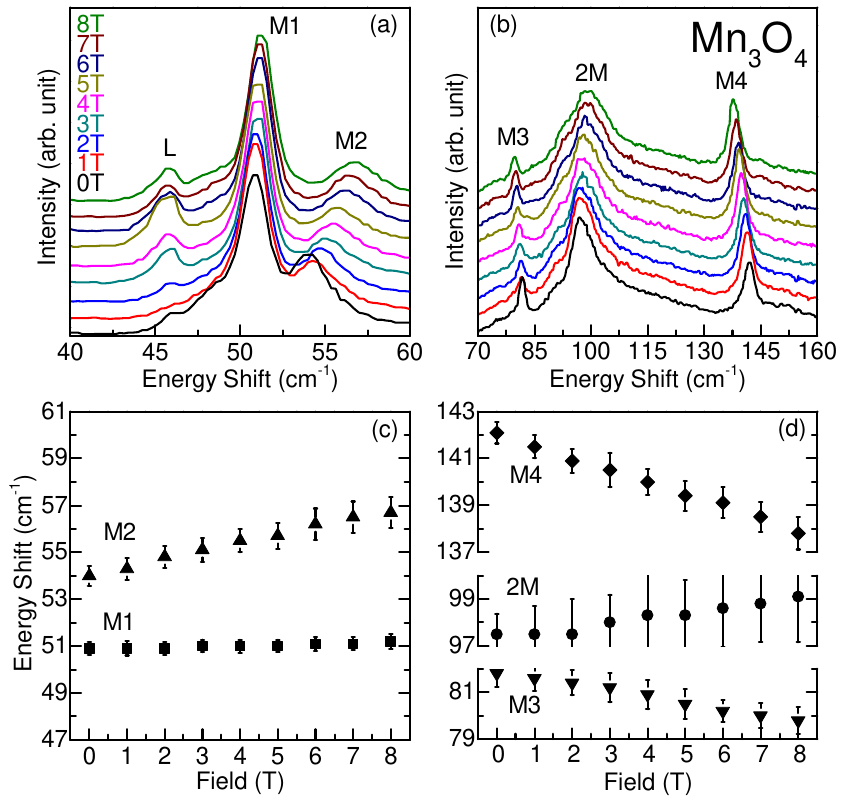}
\caption{\label{fig:Mn3O4fvH}
(color)
(a),(b) Raman scattering spectra of Mn$_{3}$O$_{4}$ at various magnetic fields in the energy ranges (a) $40-60 \textrm{ cm}^{-1}$ and (b) $70-160 \textrm{ cm}^{-1}$. 
The data have been offset for clarity. 
(c),(d) Summaries of the magnetic field dependences of the peak positions for peaks labeled M1$-$M4 and 2M, as estimated by eye.}
\end{figure}

Figure~\ref{fig:Mn3O4fvH} shows the magnetic field dependences of the one-magnon excitations M1$-$M4 in Mn$_{3}$O$_{4}$ for a magnetic field applied orthogonal to the net magnetization direction.
\footnote{Modes M1$-$M4 are also sensitive to a magnetic field applied in the ab-plane where the net magnetization resides. 
This field-dependent behavior is more complicated and will be discussed in a paper that examines the complete field dependence of the Raman spectra in Mn$_{3}$O$_{4}$ and MnV$_{2}$O$_{4}$.}
All peaks except M1 exhibit linear magnetic field dependences, with M2 increasing in energy at a rate $\Delta E/\Delta H = +0.34 \textrm{ cm}^{-1}$/T and M3 and M4 decreasing in energy at rates $\Delta E/\Delta H = -0.25 \textrm{ cm}^{-1}$/T and $\Delta E/\Delta H = -0.54 \textrm{ cm}^{-1}$/T, respectively. 
These field dependences are consistent with earlier claims\cite{Chung2008} that the one-magnon excitations identified as M3 and M4 (see Fig.~\ref{fig:Mn3O4fvT} (a)) involve spin excitations associated with the frustrated \textit{B}-site sublattice, as the Mn$^{3+}$ spins on the \textit{B}-site are reported to have an antiferromagnetic component in the direction of the applied magnetic field.\cite{Jensen1974}

We identify the broad band 2M in Mn$_{3}$O$_{4}$ (see Fig.~\ref{fig:Mn3O4fvT} (a), (c), and (e)) as two-magnon scattering. 
The temperature dependence of this response is similar to that of the one-magnon excitations in Mn$_{3}$O$_{4}$, but like the two-magnon band observed in MnV$_{2}$O$_{4}$ (peak 2M in Fig.~\ref{fig:MnV2O4fvT}), the linewidth of peak 2M in Mn$_{3}$O$_{4}$ is much broader than the one-magnon excitations. 
The broad linewidth of peak 2M is consistent with the fact that the two-magnon scattering response reflects the two-magnon density of states.\cite{Cottam1986}  
On the other hand, in contrast to the temperature dependence of the two-magnon scattering response in MnV$_{2}$O$_{4}$, the two-magnon response 2M in Mn$_{3}$O$_{4}$ exhibits a conventional temperature dependence, decreasing in energy as $T \rightarrow T_{N}$ (see Fig.~\ref{fig:Mn3O4fvT} (a), (c), and (e)).  

\section{Discussion} 

The Raman spectra of both MnV$_{2}$O$_{4}$ (Fig.~\ref{fig:MnV2O4fvT}) and Mn$_{3}$O$_{4}$ (Fig.~\ref{fig:Mn3O4fvT}) exhibit well-defined one- and two-magnon excitations that evolve below $T_{N}$ from a broad scattering continuum background. 
This continuum scattering is likely associated with scattering from short range spin correlations on the frustrated pyrochlore (\textit{B}-site) sublattice. 
This conclusion is supported by the observation of short range Mn$^{3+}$ spin correlations above $T_{N}$ in diffuse neutron scattering and specific heat measurements of Mn$_{3}$O$_{4}$.\cite{Kuriki2003}  
Notably, the spin continuum scattering near and above $T_{N}$ is significantly more prominent in Mn$_{3}$O$_{4}$ (Fig.~\ref{fig:Mn3O4fvT}) than in MnV$_{2}$O$_{4}$ (Fig.~\ref{fig:MnV2O4fvT}). 
This may reflect the much weaker \textit{B}-site interchain exchange coupling reported in Mn$_{3}$O$_{4}$ relative to MnV$_{2}$O$_{4}$,\cite{Chung2008} since spin fluctuations are expected to be enhanced in one-dimensional chain systems.\cite{Lemmens2003}

The primitive unit cells of Mn$_3$O$_4$ and MnV$_2$O$_4$ contain six magnetic ions,\cite{Jensen1974,Garlea2008} and their spin wave dispersions consist of six branches.\cite{Chung2008,Nanguneri2012} 
In Mn$_3$O$_4$, we observe four one-magnon excitations at \textbf{q} = 0 whose energies are consistent with zone center magnon energies measured previously with inelastic neutron scattering.\cite{Chung2008} 
In MnV$_2$O$_4$, on the other hand, we observe two one-magnon excitations. 
We do not necessarily expect to observe all six magnetic excitations in either material with Raman scattering, as the particular symmetry of an excitation may be inaccessible to our technique.  
Even if a magnetic excitation in a given material has a Raman-allowed symmetry, the modulation of the crystal's electronic susceptibility by this excitation may be small, leading to a weak signal. 
If this is the case, our observation of fewer one-magnon excitations in MnV$_2$O$_4$ than in Mn$_3$O$_4$ may reflect the difference between the orbital configurations of the V$^{3+}$ and Mn$^{3+}$ ions.

Significantly, while Chung \textit{et al}.\cite{Chung2008}~report a single $\textbf{q}=0$ magnon near 80~cm$^{-1}$ in MnV$_{2}$O$_{4}$, we observe two $\textbf{q}=0$ magnons at 74~cm$^{-1}$ and 81~cm$^{-1}$ in this material.
One possible explanation for this discrepancy is that the two modes we observe are associated with a slight splitting of two transverse magnon branches at $\textbf{q}=0$ --- which is not resolved by Chung \textit{et al}.\cite{Chung2008}~--- due to anisotropy effects in MnV$_{2}$O$_{4}$.
However, a more likely interpretation is that the 74~cm$^{-1}$ and 81~cm$^{-1}$ modes we observe in MnV$_{2}$O$_{4}$ are associated with a pair of one-magnon branches that were recently reported by Magee to exhibit a crossing at a momentum transfer equivalent to $\textbf{q}=0$ in MnV$_{2}$O$_{4}$.\cite{Magee2010}
This is consistent with the \emph{number} of magnon modes we observe in this energy range; importantly, however, we measure an 8~cm$^{-1}$ (1~meV) splitting between the single-magnon modes, rather than a band crossing, at $\textbf{q}=0$ in MnV$_{2}$O$_{4}$. 

The $\textbf{q}=0$ one-magnon spectrum of MnV$_{2}$O$_{4}$ we observe (see Fig.~\ref{fig:MnV2O4fvT}) should put constraints on the orbital ground state of this material. 
To clarify this point, note that there are currently several proposed orbital ordering schemes for the V$^{3+}$ ion in MnV$_{2}$O$_{4}$.\cite{Tsunetsugu2003,Nanguneri2012,Tchernyshyov2004,Khomskii2005,Chern2010,Katsufuji2013,Sarkar2009} 
Since each orbital configuration has a different electron overlap, and thus different exchange parameters, a principal test of the orbital order in MnV$_{2}$O$_{4}$ has been a comparison between exchange parameters calculated from particular orbital configurations (\textit{e.g.}, Refs.~\onlinecite{Nanguneri2012} and \onlinecite{Sarkar2009}) and exchange parameters extracted from a fit to inelastic neutron scatting data reported by Chung \textit{et al}.\cite{Chung2008} 
However, this fit does not account for our observation of two $\textbf{q}=0$ magnon excitations near 80~cm$^{-1}$ in MnV$_{2}$O$_{4}$.  
Specifically, to explain the two $\textbf{q}=0$ magnon excitations near 80~cm$^{-1}$, one of the four higher energy bands at the zone center of the fit given by Chung \textit{et al}.\cite{Chung2008} would have to be lowered by at least 8~meV. 
More recent, higher resolution neutron scattering measurements by Magee show that one of the higher energy magnon branches in MnV$_{2}$O$_{4}$ does indeed disperse downward near the zone center; however, the Magee neutron study reports a crossing of magnon branches near 80~cm$^{-1}$ at $\textbf{q}=0$ rather than the 8~cm$^{-1}$ splitting between $\textbf{q}=0$ magnon modes that we observe.\cite{Magee2010}
In short, the $\textbf{q}=0$ magnon spectrum we measure in MnV$_{2}$O$_{4}$ indicates that a revised fit to magnon dispersion data is needed to obtain improved estimates of the exchange parameters and orbital ground state in MnV$_{2}$O$_{4}$.

The two-magnon scattering responses at 178~cm$^{-1}$ in MnV$_{2}$O$_{4}$ (Fig.~\ref{fig:MnV2O4fvT}) and 98~cm$^{-1}$ in Mn$_{3}$O$_{4}$ (Fig.~\ref{fig:Mn3O4fvT}) represent the creation of pairs of spin waves with momenta $+\textbf{q}$ and $-\textbf{q}$. 
The temperature dependence of the two-magnon scattering energy $\omega_{2M}$ has been calculated by numerous authors:\cite{Solyom1971,Cottam1972,Cottam1986} 
$\omega_{2M}$ is expected to increase with decreasing temperature, following the temperature dependence of the one-magnon energy. 
The temperature dependence of the two-magnon response in MnV$_{2}$O$_{4}$ is anomalous, as the two-magnon energy \emph{decreases} with decreasing temperature (see Fig.~\ref{fig:MnV2O4fvT}~(c) and (f)). 

The two-magnon scattering intensity is proportional to the two-magnon density of states, and is thus dominated by pairs of spin waves with momenta near the edge of the Brillouin zone.
Therefore, the anomalous two-magnon temperature dependence in MnV$_2$O$_4$ must be associated with magnon interaction effects at the zone boundary.
Strong spin-lattice coupling associated with the zone boundary magnons in MnV$_{2}$O$_{4}$ is the simplest and most likely explanation for the anomalous two-magnon temperature dependence, particularly given the evidence for strong spin-lattice coupling apparent in the closely spaced magnetic and structural phase transitions of MnV$_{2}$O$_{4}$.
A similar temperature dependence of the two-magnon band in the spin-dimer system TlCuCl$_{3}$ was also observed and attributed to strong spin-phonon coupling.\cite{Choi2003} 

Notably, there is no evidence for anomalies associated with either of the $\textbf{q}=0$ one-magnon peaks M1 and M2 in MnV$_2$O$_4$.
These $\textbf{q}=0$ magnon modes are well separated in energy from --- and hence are not expected to mix with --- the $\textbf{q}=0$ phonon modes of MnV$_2$O$_4$.\cite{Myung-Whun2012,Takubo2011}
Strong spin-lattice coupling should be more pronounced near the Brillouin zone boundary in this material, where the dispersed acoustic phonons (or possibly softened optical phonons) are more likely to overlap in energy with magnons.

Our Raman results motivate a more conclusive investigation of possible mixing of the spin and lattice degrees of freedom at the zone boundary in MnV$_{2}$O$_{4}$, particularly an inelastic neutron scattering search for anomalies in the temperature dependence of zone boundary magnons and phonons in the $9-12$~meV range. 
For example, neutron scattering studies of the manganese perovskites revealed spin wave softening and linewidth broadening of the zone boundary magnons caused by spin-phonon coupling.\cite{Dai2000}

In Mn$_3$O$_4$, we did not find a similar anomalous temperature dependence for any of the magnetic excitations we were able to observe.
Spin-lattice coupling is known to have significant effects in Mn$_3$O$_4$,\cite{Chung2013,Suzuki2008,Tackett2007,Suzuki2006,Kim2011,Kim2010} which raises the question of why we do not observe magnon anomalies in Mn$_3$O$_4$ similar to that observed in MnV$_2$O$_4$.
A likely reason for this discrepancy is simply that magnon-phonon coupling in Mn$_3$O$_4$ occurs in a region of the Brillouin zone that is inaccessible to Raman scattering.
Temperature dependent neutron scattering studies, which would allow a study of magnon and phonon anomalies throughout the Brillouin zone of Mn$_3$O$_4$, are needed to provide a more comprehensive search for magnon anomalies associated with strong spin-lattice coupling.

\section{Summary}

In summary, we have measured the temperature and magnetic field dependences of one- and two-magnon excitations in the magnetically frustrated spinels MnV$_{2}$O$_{4}$ and Mn$_{3}$O$_{4}$.  
Both materials exhibit diffusive spin scattering above $T_{N}$, which we associate with incoherent spin fluctuations on the \textit{B}-site sublattice. 
Below $T_{N}$, one- and two-magnon excitations develop in both MnV$_{2}$O$_{4}$ and Mn$_{3}$O$_{4}$.
In Mn$_{3}$O$_{4}$, the zone center magnon spectrum we observe matches well with previous neutron scattering results and spin wave calculations.  
However, in MnV$_{2}$O$_{4}$, we observe a pair of zone center magnon modes near 80~cm$^{-1}$, one of which is not predicted by either current spin wave calculations or the fit to inelastic neutron scattering data of Chung \textit{et al}.\cite{Chung2008}
The $\textbf{q}=0$ magnon results on MnV$_{2}$O$_{4}$ presented here, along with more recent neutron scattering measurements of the magnon branches in MnV$_{2}$O$_{4}$ by Magee,\cite{Magee2010} indicate that a revised fit to the magnon dispersion data for MnV$_{2}$O$_{4}$ is needed to obtain improved estimates of the exchange parameters and orbital ground state in MnV$_{2}$O$_{4}$.  
Finally, we show that the two-magnon energy of Mn$_{3}$O$_{4}$ exhibits a conventional temperature dependence, but that the two-magnon energy of MnV$_{2}$O$_{4}$ shows an anomalous softening with decreasing temperature below $T_{N}$.
We propose that this softening reflects strong coupling between vibrational and magnetic excitations near the zone boundary in MnV$_{2}$O$_{4}$, and we suggest that high-resolution neutron scattering studies of zone boundary magnons and/or phonons are needed to further elucidate the nature of this strong spin-phonon coupling.

\begin{acknowledgments}
We thank S.~E.~Nagler for valuable conversations.
Research was supported by the U.S. Department of Energy, Office of Basic Energy Sciences, Division of Materials Sciences and Engineering under Award DE-FG02-07ER46453. 
T. Byrum was partially supported by the National Science Foundation Graduate Research Fellowship Program under Grant Number DGE-1144245. 
\end{acknowledgments}

\end{document}